\documentclass[11pt]{article}
\usepackage{epic,latexsym,amssymb}
\usepackage{color}
\usepackage{tikz}

\newtheorem{theorem}{Theorem}

\newtheorem{prop}{Proposition}
\newtheorem{conj}{Conjecture}

\newtheorem{claim}{Claim}
\newtheorem{subclaim}{Claim}[claim]

\newcommand{\QED}{$\Box$}
\newcommand{\smallqed}{{\tiny ($\Box$)}}

\newcommand{\coro}{{\rm cor}}
\newcommand{\modo}{{\rm mod \,}}

\newcommand{\cFc}{\mathcal{F}_{\mathrm{cubic}}}
\newcommand{\cGc}{\mathcal{G}_{\mathrm{cubic}}}
\newcommand{\cHc}{\mathcal{H}_{\mathrm{cubic}}}

\newcommand{\proof}{\noindent\textbf{Proof.} }
\newcommand{\2}{ \vspace{0.2cm} }
\newcommand{\1}{ \vspace{0.1cm} }

\let\oldenumerate\enumerate
\renewcommand{\enumerate}{
  \oldenumerate
  \setlength{\itemsep}{0pt}
  \setlength{\parskip}{0pt}
  \setlength{\parsep}{0pt}
}

\begin{document}

\title{Independent Domination in Subcubic Graphs}

\author{$^1$A. Akbari, $^1$S. Akbari, $^1$A. Doosthosseini, \\ $^1$Z. Hadizadeh, $^2$Michael A. Henning, and $^1$A. Naraghi
\\ \\
$^1$Department of Mathematical Sciences \\
Sharif University of Technology\\
Tehran, Iran \\
\\
$^2$Department of Mathematics and Applied Mathematics \\
University of Johannesburg \\
Johannesburg, South Africa}

\footnotetext{E-mail Addresses:
\tt{ami.akbari99@student.sharif.edu},
\tt{s\_akbari@sharif.edu}, \tt{a.doosth77@student.sharif.edu},
\tt{zahra.hadizadeh78@student.sharif.edu},
\tt{mahenning@uj.ac.za}, \tt{a.naraghi2015@student.sharif.edu}}

\date{}
\maketitle

\begin{abstract}
A set $S$ of vertices in a graph $G$ is a dominating set if every vertex not in $S$ is adjacent to a vertex in~$S$. If, in addition, $S$ is an independent set, then $S$ is an independent dominating set. The independent domination number $i(G)$ of $G$ is the minimum cardinality of an independent dominating set in $G$. In 2013 Goddard and Henning [Discrete Math 313 (2013), 839--854] conjectured that if $G$ is a connected cubic graph of order~$n$, then $i(G) \le \frac{3}{8}n$, except if $G$ is the complete bipartite graph $K_{3,3}$ or the $5$-prism $C_5 \, \Box \, K_2$. Further they construct two infinite families of connected cubic graphs with independent domination three-eighths their order. They remark that perhaps it is even true that for $n > 10$ these two families are only families for which equality holds. In this paper, we provide a new family of connected cubic graphs $G$ of order $n$ such that $i(G) = \frac{3}{8}n$. We also show that if $G$ is a subcubic graph of order $n$ with no isolated vertex, then $i(G) \le \frac{1}{2}n$, and we characterize the graphs achieving equality in this bound.
\end{abstract}


\newpage
\section{Introduction}

A set $S$ of vertices in a graph $G$ is a \emph{dominating set} if every vertex not in $S$ is adjacent to a vertex in~$S$. If, in addition, $S$ is an independent set, then $S$ is an \emph{independent dominating set}, abbreviated ID-set.  The \emph{independent domination number}, denoted $i(G)$, of $G$ is the minimum cardinality  of an ID-set in $G$. The concept of independent domination number of graphs is studied extensively in the literature, for example see~\cite{AbHe18,BrHe19,DoHeMoSo15,Fa,Furuya,GoHeLySo12,GoLy12,Ha95,Ha07,HeLoRa14,Ko93,LSS99,Ly14,Ly15,OWe10,RaVo13,SoHe13,WaWe}.
A survey on independent domination in graphs can be found in~\cite{GoHe13}.

For notation and graph theory terminology we generally follow~\cite{HeYe_book}. The \emph{order} of a graph $G$ with vertex set $V(G)$ and edge set $E(G)$ is denoted by $n(G) = |V(G)|$ and its \emph{size} by $m(G) = |E(G)|$. Two vertices are \emph{neighbors} in $G$ if they are adjacent. The \emph{open neighborhood} of a vertex $v$ in $G$ is the set of neighbors of $v$, denoted $N_G(v)$. Thus, $N_G(v) = \{u \in V(G) \mid uv \in E(G)\}$. The \emph{closed neighborhood} of $v$ is the set  $N_G[v] = N_G(v) \cup \{v\}$. The \emph{degree} of a vertex $v$ in $G$ is denoted $d_G(v) = |N_G(v)|$. We denote the minimum and maximum degrees among the vertices of $G$ by $\delta(G)$ and $\Delta(G)$, respectively. A \emph{cubic graph} is a graph in which every vertex has degree~$3$, while a \emph{subcubic graph} is a graph with maximum degree at most~$3$.

Further, the subgraph obtained from $G$ by deleting all vertices in $S$ and all edges incident with vertices in $S$ is denoted by $G - S$. If $S = \{v\}$, we simply denote $G - \{v\}$ by $G - v$. A \emph{leaf} of a graph $G$ is a vertex of degree~$1$ in $G$, while a \emph{support vertex} of $G$ is a vertex adjacent to a leaf. A \emph{star} is the graph $K_{1,k}$, where $k \ge 1$; that is, a star is a tree with at most one vertex that is not a leaf. A \emph{double star} is a tree with exactly two (adjacent) non-leaf vertices. Further if one of these vertices is adjacent to $r$ leaves and the other to $s$ leaves, then we denote the double star by $S(r,s)$.
We denote the path and cycle on $n$ vertices by $P_n$ and $C_n$, respectively, and we denote a complete bipartite with partite sets of cardinalities~$n$ and~$m$ by $K_{n,m}$. The \emph{corona} $\coro(G)$ of a graph $G$, also denoted $G \circ P_1$ in the literature, is the graph obtained from $G$ by adding a pendant edge to each vertex of $G$. For $k \ge 1$ an integer, we use the standard notation $[k] = \{1,\ldots,k\}$ and $[k]_0 = \{0,1,\ldots,k\}$.

\section{Motivation and Known Results}

As remarked in~\cite{GoHe13}, since every bipartite graph is the union of two independent sets, each of which dominates the other, we have the following well-known bound on the independent domination number of a bipartite graph.

\begin{prop}
\label{p:bipartite}
If $G$ is a bipartite graph with no isolated vertices of order~$n$, then $i(G) \le \frac{1}{2}n$.
\end{prop}

As noted in~\cite{GoHe13}, the bound in Proposition~\ref{p:bipartite} is sharp as may be seen by taking $G = K_{k,k}$ for any $k \ge 1$. In particular, if $G = K_{k,k}$ and $k \in [3]$, then $G$ is a connected subcubic graph of order~$n = 2k$ such that $i(G) = \frac{1}{2}n$.

\begin{figure}[htb]
\begin{center}
\begin{tikzpicture}
[thick,scale=0.75,
	vertex/.style={circle,draw,inner sep=0pt,minimum size=1.5mm},
	vertexlabel/.style={circle,draw=white,inner sep=0pt,minimum size=0mm}]
\def \irad {.7}
\def \orad {1.5}
\def \ydist {1}
\coordinate (A) at (0,0);
\draw (A)
{	
	node [fill=black] [vertex] at (-1.5,-\ydist) (x1){}
	node [fill=black] [vertex] at (0,-\ydist) (x2){}
	node [fill=black] [vertex] at (1.5,-\ydist) (x3){}
    node [fill=black] [vertex] at (-1.5,\ydist) (y1){}
	node [fill=black] [vertex] at (0,\ydist) (y2){}
	node [fill=black] [vertex] at (1.5,\ydist) (y3){}
	(y1)--(x1) (y1)--(x2) (y1)--(x3)
(y2)--(x1) (y2)--(x2) (y2)--(x3)
(y3)--(x1) (y3)--(x2) (y3)--(x3)
	(A) +(0,-2) node [rectangle, draw=white!0, fill=white!100]{\small (a) $K_{3,3}$}
};
\coordinate (B) at (6,0);

\foreach \i in {0,1,2,3,4}
{
	\path (B) node [fill=black] [vertex] at +(90 - \i*360/5:\irad) (x\i){};
	\path (B) node [fill=black] [vertex] at +(90 - \i*360/5:\orad) (y\i){};
	\draw (x\i)--(y\i);
}
	\draw (x0)--(x1)--(x2)--(x3)--(x4)--(x0);
	\draw (y0)--(y1)--(y2)--(y3)--(y4)--(y0);

\draw (B)
{	
	+(0,-2) node [rectangle, draw=white!0, fill=white!100]{\small (b) $C_5 \, \Box \, K_2$}
};
\end{tikzpicture}\end{center}
\vskip -0.7 cm \caption{The graphs $K_{3,3}$ and $C_5 \, \Box \, K_2$.} \label{f:small graphs}
\end{figure}
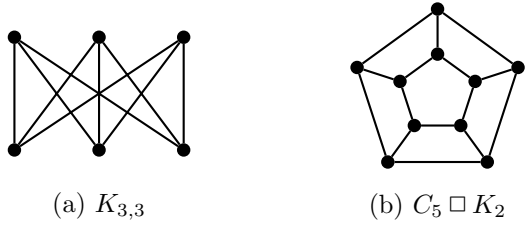

It remains an open problem to determine best possible upper bounds on the independent domination number of a connected cubic graph in terms of its orders. Lam, Shiu, and Sun~\cite{LSS99} proved that if $G$ is a connected cubic graph of order~$n$ different from $K_{3,3}$, then $i(G) \le \frac{2}{5}n$, where the graph $K_{3,3}$ is given in Figure~\ref{f:small graphs}(a). This bound is achieved by the $5$-prism $C_5 \, \Box \, K_2$ which is illustrated in Figure~\ref{f:small graphs}(b).

Goddard and Henning~\cite{GoHe13} posed the conjecture that the $\frac{2}{5}n$ bound on the independent domination number can be improved if we forbid the exceptional graphs $K_{3,3}$ and $C_5 \, \Box \, K_2$.

\begin{conj}{\rm (\cite{GoHe13})}
\label{c:cubic3n8I}
If $G \notin \{K_{3,3}, C_5 \, \Box \, K_2\}$
is a connected, cubic graph of order~$n$, then $i(G) \le \frac{3}{8}n$.
\end{conj}

Dorbec, Henning, Montassier, and Southey~\cite{DoHeMoSo15} proved Conjecture~\ref{c:cubic3n8I} in the case when there is no subgraph isomorphic to $K_{2,3}$.  In general, however, Conjecture~\ref{c:cubic3n8I} remains unresolved.

\begin{figure}[htb]
\begin{center}
\begin{tikzpicture}[thick,scale=.65]%
\tikzstyle{every node}=[circle, draw, fill=black!100, inner sep=0pt,minimum width=.17cm]
  \draw [densely dashed] (0,0) {
    +(0,2.5) -- +(0,3.25) +(2,2.5) -- +(2,3.25)};
  \draw [rounded corners=8pt] (0,0) {
    +(0,0) -- +(-.5,0) -- +(-.5,5.75) -- +(0,5.75)
    +(2,0) -- +(2.5,0) -- +(2.5,5.75) -- +(2,5.75)};
  \draw (0,0) {
    +(0,0) -- +(2,.75) +(0,.75) -- +(2,0)
    +(0,1.5) -- +(2,1.5) +(0,2.25) -- +(2,2.25)};
  \draw (0,3.5) {
    +(0,0) -- +(2,.75) +(0,.75) -- +(2,0)
    +(0,1.5) -- +(2,1.5) +(0,2.25) -- +(2,2.25)};
  \draw (0,0) {
    +(0,0.00) node{} -- +(0,0.75) node{} -- +(0,1.5) node{} -- +(0,2.25) node{} -- +(0,2.5)
    +(0,3.25) -- +(0,3.5) node{} -- +(0,4.25) node{} -- +(0,5.0) node{} -- +(0,5.75) node{}};
  \draw (2,0) {
    +(0,0.00) node{} -- +(0,0.75) node{} -- +(0,1.5) node{} -- +(0,2.25) node{} -- +(0,2.5)
    +(0,3.25) -- +(0,3.5) node{} -- +(0,4.25) node{} -- +(0,5.0) node{} -- +(0,5.75) node{}};
  \node[rectangle, draw=white!0, fill=white!100] at (1,-.75) {(a) $G$};
  \draw [densely dashed] (6,0) {
    +(0,2.5) -- +(0,3.25)};
  \draw [rounded corners=8pt] (6,0) {
    +(0,0) -- +(-.5,0) -- +(-.5,5.75) -- +(0,5.75)};
  \draw (6,0) {
    +(2,2) -- +(4,.5)        +(2,2) -- +(4,1.5)
    +(2,1) -- +(4,.5)        +(2,1) -- +(4,1.5)
    +(2,0) -- +(4,.5) node{} +(2,0) -- +(4,1.5) node{}};
  \draw (6,3.75) {
    +(2,2) -- +(4,.5)        +(2,2) -- +(4,1.5)
    +(2,1) -- +(4,.5)        +(2,1) -- +(4,1.5)
    +(2,0) -- +(4,.5) node{} +(2,0) -- +(4,1.5) node{}};
  \draw (6,0) {
    +(0,0) -- +(0,2.5) +(0,3.25) -- +(0,5.75)};
  \draw (6,0) {
    +(0,0) node{} -- +(2,0) node{}
    +(0,1) node{} -- +(2,1) node{}
    +(0,2) node{} -- +(2,2) node{}
    +(0,3.75) node{} -- +(2,3.75) node{}
    +(0,4.75) node{} -- +(2,4.75) node{}
    +(0,5.75) node{} -- +(2,5.75) node{}};
  \node[rectangle, draw=white!0, fill=white!100] at (7,-.75) {(b) $H$};
\end{tikzpicture}
\end{center}
\vskip -0.6 cm \caption{Graphs $G \in \cGc$ and $H \in \cHc$ of order~$n$ with $i(G) = i(H) =  \frac{3}{8}n$.} \label{GH}
\end{figure}
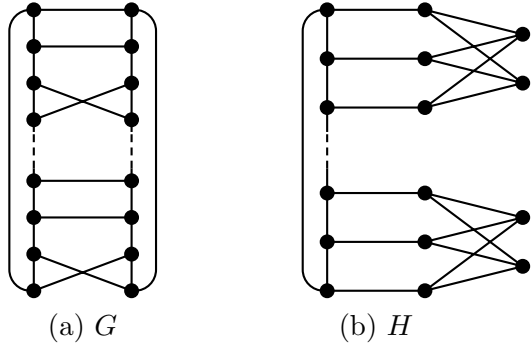

Goddard and Henning~\cite{GoHe13}  constructed two infinite families $\cGc$ and $\cHc$ of connected cubic graphs with independent domination number three-eighths their orders  as follows. For $k \ge 1$, a graph in the family $\cGc$ is constructed by taking two copies of the cycle $C_{4k}$ with respective vertex sequences $a_1b_1c_1d_1 \ldots a_kb_kc_kd_k$ and $w_1x_1y_1z_1 \ldots w_kx_ky_kz_k$, and joining $a_i$ to $w_i$, $b_i$ to $x_i$, $c_i$ to $z_i$, and $d_i$ to $y_i$ for each $i \in [k]$.
For $\ell \ge 1$, a graph in the family $\cHc$ is constructed by taking a copy of a cycle $C_{3\ell}$ with vertex sequence $a_1b_1c_1 \ldots a_{\ell} b_{\ell} c_{\ell}$, and for each $i \in [\ell]$, adding the vertices $\{w_i,x_i,y_i,z^1_i,z^2_i\}$, and joining $a_i$ to $w_i$, $b_i$ to $x_i$, and $c_i$ to $y_i$, and further for each $j \in [2]$, joining $z^j_i$ to each of the vertices $w_i$, $x_i$, and $y_i$.
Graphs in the families $\cGc$ and $\cHc$ are illustrated in Figure~\ref{GH}(a) and~\ref{GH}(b), respectively.

\begin{theorem}{\rm (\cite{GoHe13})}
\label{t:tight1}
If $G \in \cGc \cup \cHc$ has order~$n$, then $i(G) = \frac{3}{8}n$.
\end{theorem}

It is remarked in~\cite{GoHe13} that ``Perhaps even more than Conjecture~\ref{c:cubic3n8I} is true, in that the only extremal graphs are those in $\cGc \cup \cHc$. We have confirmed by computer search that this is true when $n \le 20$."

We remark that several papers, see for example~\cite{AbHe18,BrHe19,DoHeMoSo15,HeLoRa14,SoHe13}, in which upper bounds are obtained on the independent domination number of cubic graphs present more general results on subcubic graphs.

\section{Main Results}

In this paper we have two immediate aims. Our first aim is to provide a new family of connected cubic graphs, different from the families $\cGc$ and $\cHc$, such that every graph $G$ of order~$n$ in the family satisfies $i(G) = \frac{3}{8}n$. We shall prove the following result, where $\cFc$ is the family of connected cubic graphs constructed in
Section~\ref{S:main1}.

\begin{theorem}
\label{t:main1}
If $G \in \cFc$ has order~$n$, then $n \ge 16$ and $n \equiv 0 \, (\modo \, 8)$ and $i(G) = \frac{3}{8}n$.
\end{theorem}

\begin{figure}[htb]
\begin{center}
\begin{tikzpicture}[scale=.8,style=thick,x=1cm,y=1cm]
\def\vr{2.5pt} 
\path (0,0.75) coordinate (a1);
\path (1,0) coordinate (a2);
\path (1,0.75) coordinate (a3);
\path (1,1.5) coordinate (a4);
\path (2,0.75) coordinate (a5);
\path (3,0.75) coordinate (a6);
%
\draw (a2) -- (a1);
\draw (a2) -- (a3);
\draw (a2) -- (a5);
\draw (a4) -- (a1);
\draw (a4) -- (a3);
\draw (a4) -- (a5);
\draw (a5) -- (a6);
\draw (a1) [fill=black] circle (\vr);
\draw (a2) [fill=black] circle (\vr);
\draw (a3) [fill=black] circle (\vr);
\draw (a4) [fill=black] circle (\vr);
\draw (a5) [fill=black] circle (\vr);
\draw (a6) [fill=black] circle (\vr);
\draw (1.5,-0.75) node {{\small (a) $G_3$}};
\path (5,0) coordinate (b1);
\path (5,1.5) coordinate (b2);
\path (6,0.75) coordinate (b3);
\path (7,0.75) coordinate (b4);
\path (8,0) coordinate (b5);
\path (8,1.5) coordinate (b6);
%
\draw (b3) -- (b1);
\draw (b3) -- (b2);
\draw (b3) -- (b4);
\draw (b4) -- (b5);
\draw (b4) -- (b6);
\draw (b1) [fill=black] circle (\vr);
\draw (b2) [fill=black] circle (\vr);
\draw (b3) [fill=black] circle (\vr);
\draw (b4) [fill=black] circle (\vr);
\draw (b5) [fill=black] circle (\vr);
\draw (b6) [fill=black] circle (\vr);
\draw (6.5,-0.75) node {{\small (b) $G_4$}};
\path (10,0) coordinate (c1);
\path (10,1.5) coordinate (c2);
\path (11,0) coordinate (c3);
\path (11,1.5) coordinate (c4);
\path (12,0) coordinate (c5);
\path (12,1.5) coordinate (c6);
%
\draw (c1) -- (c2);
\draw (c1) -- (c3);
\draw (c2) -- (c4);
\draw (c3) -- (c5);
\draw (c3) -- (c4);
\draw (c4) -- (c6);
\draw (c1) [fill=black] circle (\vr);
\draw (c2) [fill=black] circle (\vr);
\draw (c3) [fill=black] circle (\vr);
\draw (c4) [fill=black] circle (\vr);
\draw (c5) [fill=black] circle (\vr);
\draw (c6) [fill=black] circle (\vr);
\draw (11,-0.75) node {{\small (c) $G_5$}};
\end{tikzpicture}
\end{center}
\vskip -0.7 cm
\caption{The graphs $G_3$, $G_4$ and $G_5$} \label{f:special}
\end{figure}
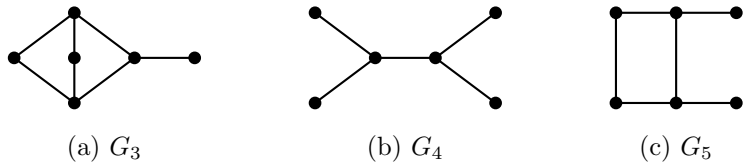

Our second aim is to provide a tight upper bound on the independent domination number of a subcubic graph, and to characterize the graphs achieving equality in this bound. Let $G_1 \cong K_{2,2}$ and $G_2 \cong K_{3,3}$, and let $G_3, G_4, G_5$ be the three graphs shown in Figure~\ref{f:special}.

We shall prove the following result, a proof of which is presented in Section~\ref{S:main1}.

\begin{theorem}
\label{t:main2}
If $G$ is a subcubic graph of order~$n$ with no isolated vertex, then $i(G) \le \frac{1}{2}n$, with equality if and only if the following holds. \\ [-28pt]
\begin{enumerate}
\item $G \in \{G_1,G_2,G_3,G_4,G_5\}$.
\item $n = 2k$ for some $k \ge 1$ and $G = \coro(P_k)$.
\item $n = 2k$ for some $k \ge 3$ and $G = \coro(C_k)$.
\end{enumerate}
\end{theorem}

\section{Proof of Theorem~\ref{t:main1}}
\label{S:main1}

\begin{figure}[htb]
\begin{center}
\begin{tikzpicture}[scale=.8,style=thick,x=1cm,y=1cm]
\def\vr{2.5pt} 
\path (0.5,0) coordinate (a1);
\path (1.5,0) coordinate (a2);
\path (0,1) coordinate (a3);
\path (0,2) coordinate (a6);
\path (1,1) coordinate (a4);
\path (1,2) coordinate (a7);
\path (2,1) coordinate (a5);
\path (2,2) coordinate (a8);
%
\draw (a1) -- (a3);
\draw (a1) -- (a4);
\draw (a1) -- (a5);
\draw (a2) -- (a3);
\draw (a2) -- (a4);
\draw (a2) -- (a5);
\draw (a6) -- (a3);
\draw (a7) -- (a4);
\draw (a8) -- (a5);
\draw (a7) -- (a8);
\draw (a1) [fill=black] circle (\vr);
\draw (a2) [fill=black] circle (\vr);
\draw (a3) [fill=black] circle (\vr);
\draw (a4) [fill=black] circle (\vr);
\draw (a5) [fill=black] circle (\vr);
\draw (a6) [fill=black] circle (\vr);
\draw (a7) [fill=black] circle (\vr);
\draw (a8) [fill=black] circle (\vr);
\draw (1,-0.75) node {{\small (a) $X$}};
\path (5.5,0) coordinate (b1);
\path (6.5,0) coordinate (b2);
\path (5,1) coordinate (b3);
\path (5,2) coordinate (b6);
\path (6,1) coordinate (b4);
\path (6,2) coordinate (b7);
\path (7,1) coordinate (b5);
\path (7,2) coordinate (b8);
%
\draw (b1) -- (b3);
\draw (b1) -- (b4);
\draw (b1) -- (b5);
\draw (b2) -- (b3);
\draw (b2) -- (b4);
\draw (b2) -- (b5);
\draw (b6) -- (b3);
\draw (b7) -- (b4);
\draw (b8) -- (b5);
\draw (b6) -- (b7);
\draw (b7) -- (b8);
\draw (b1) [fill=black] circle (\vr);
\draw (b2) [fill=black] circle (\vr);
\draw (b3) [fill=black] circle (\vr);
\draw (b4) [fill=black] circle (\vr);
\draw (b5) [fill=black] circle (\vr);
\draw (b6) [fill=black] circle (\vr);
\draw (b7) [fill=black] circle (\vr);
\draw (b8) [fill=black] circle (\vr);
\draw (6,-0.75) node {{\small (b) $Y$}};
\end{tikzpicture}
\end{center}
\vskip -0.7 cm
\caption{The graphs $X$ and $Y$} \label{f:XY}
\end{figure}
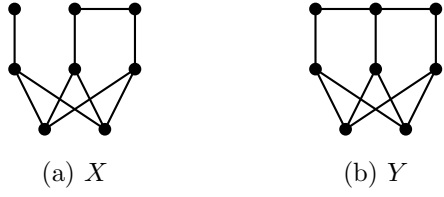

Let $X$ and $Y$ be the graphs shown in Figure~\ref{f:XY}(a) and~\ref{f:XY}(b), respectively.

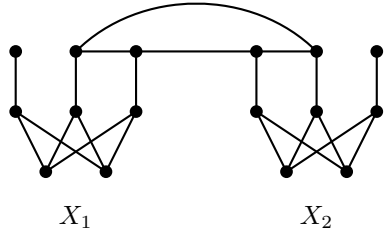
\begin{figure}[htb]
\begin{center}
\begin{tikzpicture}[scale=.8,style=thick,x=1cm,y=1cm]
\def\vr{2.5pt} 
\path (0.5,0) coordinate (a1);
\path (1.5,0) coordinate (a2);
\path (0,1) coordinate (a3);
\path (0,2) coordinate (a6);
\path (1,1) coordinate (a4);
\path (1,2) coordinate (a7);
\path (2,1) coordinate (a5);
\path (2,2) coordinate (a8);
%
\draw (a1) -- (a3);
\draw (a1) -- (a4);
\draw (a1) -- (a5);
\draw (a2) -- (a3);
\draw (a2) -- (a4);
\draw (a2) -- (a5);
\draw (a6) -- (a3);
\draw (a7) -- (a4);
\draw (a8) -- (a5);
\draw (a7) -- (a8);
\draw (a1) [fill=black] circle (\vr);
\draw (a2) [fill=black] circle (\vr);
\draw (a3) [fill=black] circle (\vr);
\draw (a4) [fill=black] circle (\vr);
\draw (a5) [fill=black] circle (\vr);
\draw (a6) [fill=black] circle (\vr);
\draw (a7) [fill=black] circle (\vr);
\draw (a8) [fill=black] circle (\vr);
\draw (1,-0.75) node {{\small $X_1$}};
\path (4.5,0) coordinate (b1);
\path (5.5,0) coordinate (b2);
\path (4,1) coordinate (b3);
\path (4,2) coordinate (b6);
\path (5,1) coordinate (b4);
\path (5,2) coordinate (b7);
\path (6,1) coordinate (b5);
\path (6,2) coordinate (b8);
%
\draw (b1) -- (b3);
\draw (b1) -- (b4);
\draw (b1) -- (b5);
\draw (b2) -- (b3);
\draw (b2) -- (b4);
\draw (b2) -- (b5);
\draw (b6) -- (b3);
\draw (b7) -- (b4);
\draw (b8) -- (b5);
\draw (b6) -- (b7);
\draw (b1) [fill=black] circle (\vr);
\draw (b2) [fill=black] circle (\vr);
\draw (b3) [fill=black] circle (\vr);
\draw (b4) [fill=black] circle (\vr);
\draw (b5) [fill=black] circle (\vr);
\draw (b6) [fill=black] circle (\vr);
\draw (b7) [fill=black] circle (\vr);
\draw (b8) [fill=black] circle (\vr);
\draw (5,-0.75) node {{\small $X_2$}};
\draw (a7) to[out=45,in=135, distance=1.5cm ] (b7);
\draw (a8) -- (b6);
\end{tikzpicture}
\end{center}
\vskip -0.7 cm
\caption{Joining of the pairs $X_1$ and $X_2$} \label{f:Xadd}
\end{figure}

Let $\cFc$ be the family of graphs constructed as follows. A graph $G$ in the family $\cFc$ is constructed as follows. We start by taking a cycle $C \colon v_1v_2 \ldots v_kv_1$ on $k \ge 2$ (for $k = 2$ we mean two vertices adjacent with two different edges) vertices and coloring every vertex on the cycle $C$, red or blue in such a way that the number of red vertices is even. We then replace each red vertex on $C$ with a copy of $X$, and each blue vertex on $C$ with a copy of $Y$. (In the case $k=2$ we only replace each vertex by a copy of $Y$.) We call each resulting copy of $X$ and $Y$ an $X$-\emph{copy} and $Y$-\emph{copy} of $G$, respectively. Let $G_i$ be the $X$-copy or $Y$-copy associated with the vertex $v_i$ on the cycle $C$ for each $i \in [k]$. Thus if $v_i$ is colored red, then $G_i \cong X$, while if $v_i$ is colored blue, then $G_i \cong Y$ for $i \in [k]$. We note that there are an even number of $X$-copies in $G$. Next we partition these $X$-copies into pairs. For each resulting pair $\{X_1,X_2\}$ where $X_i \cong X$ for $i\in [2]$, we add two edges as follows: If $v_{i1}$ and $v_{i2}$ denote the two (adjacent) vertices of degree~$2$ in $X_i$ for $i \in [2]$, then we add the edges $v_{11}v_{21}$ and  $v_{12}v_{22}$ as illustrated in Figure~\ref{f:Xadd}.

\begin{figure}[h!]
\centering
\includegraphics[scale=0.4]{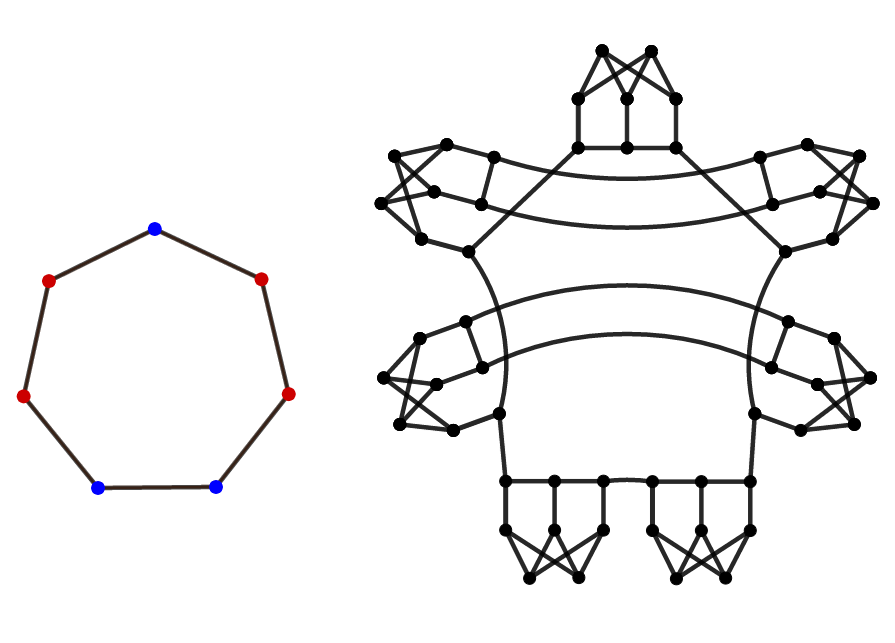}
\caption{A graph $G$ in the family $\cFc$ constructed from a colored $7$-cycle}
\label{fig:family}
\end{figure}

We note that each $X$-copy of $G$ contains a vertex of degree~$1$ and each $Y$-copy of $G$ contains two vertices of degree~$2$. We now complete the construction of the graph $G$ as follows. Consider the subgraphs $G_i$ and $G_{i+1}$ where addition is taken modulo~$k$ and where $i \in [k]$. If $G_i$ is an $X$-copy, then let $x_i$ denote the vertex of degree~$1$ in $G_i$, while if $G_i$ is a $Y$-copy, then let $y_{i}^1$ and $y_{i}^2$ denote the two vertices of degree~$2$ in $G_i$. If both $G_i$ and $G_{i+1}$ are $X$-copies, then add the edge $x_1x_2$. If both $G_i$ and $G_{i+1}$ are $Y$-copies, then add the edge $y_{i}^2y_{i+1}^1$. If $G_i$ is an $X$-copy and $G_{i+1}$ is a $Y$-copy, then add the edge $x_1y_{i+1}^1$. If $G_i$ is a $Y$-copy and $G_{i+1}$ is an $X$-copy, then add the edge $y_{i}^2x_{i+1}$. We do this for each $i \in [k]$, and let $G$ denote the resulting graph. An example of a graph $G$ in the family $\cFc$ constructed from a colored $7$-cycle (here $k = 7$) with four red vertices and three blue vertices is illustrated in Figure~\ref{fig:family}.
	
We are now in a position to prove Theorem~\ref{t:main1}. Recall its statement.

\noindent \textbf{Theorem~\ref{t:main1}}. \emph{If $G \in \cFc$ has order~$n$, then $n \ge 16$ and $n \equiv 0 \, (\modo \, 8)$ and $i(G) = \frac{3}{8}n$.
}

\noindent
\proof If $G \in \cFc$ has order~$n$, then by construction $G$ is obtained from a $k$-cycle for some $k \ge 2$ by replacing each vertex with a copy of $X$ or $Y$ and adding certain edges to produce a connected cubic graph. Since each copy of $X$ and $Y$ has order~$8$, we note that $n = 8k$. Thus, $n \ge 16$ and $n \equiv 0 \, (\modo \, 8)$. Next we show that $i(G) = \frac{3}{8}n$.

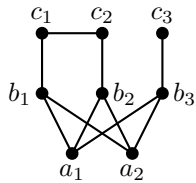
\begin{figure}[htb]
\begin{center}
\begin{tikzpicture}[scale=.8,style=thick,x=1cm,y=1cm]
\def\vr{2.5pt} 
\path (0.5,0) coordinate (a1);
\path (1.5,0) coordinate (a2);
\path (0,1) coordinate (a3);
\path (0,2) coordinate (a6);
\path (1,1) coordinate (a4);
\path (1,2) coordinate (a7);
\path (2,1) coordinate (a5);
\path (2,2) coordinate (a8);
%
\draw (a1) -- (a3);
\draw (a1) -- (a4);
\draw (a1) -- (a5);
\draw (a2) -- (a3);
\draw (a2) -- (a4);
\draw (a2) -- (a5);
\draw (a6) -- (a3);
\draw (a7) -- (a4);
\draw (a8) -- (a5);
\draw (a7) -- (a6);
\draw (a1) [fill=black] circle (\vr);
\draw (a2) [fill=black] circle (\vr);
\draw (a3) [fill=black] circle (\vr);
\draw (a4) [fill=black] circle (\vr);
\draw (a5) [fill=black] circle (\vr);
\draw (a6) [fill=black] circle (\vr);
\draw (a7) [fill=black] circle (\vr);
\draw (a8) [fill=black] circle (\vr);
\draw[anchor = north] (a1) node {{\small $a_1$}};
\draw[anchor = north] (a2) node {{\small $a_2$}};
\draw[anchor = east] (a3) node {{\small $b_1$}};
\draw[anchor = west] (a4) node {{\small $b_2$}};
\draw[anchor = west] (a5) node {{\small $b_3$}};
\draw[anchor = south] (a6) node {{\small $c_1$}};
\draw[anchor = south] (a7) node {{\small $c_2$}};
\draw[anchor = south] (a8) node {{\small $c_3$}};
\end{tikzpicture}
\end{center}
\vskip -0.7 cm
\caption{An $X$-copy in $G$} \label{f:Xcopy}
\end{figure}

Let $S$ be an arbitrary ID-set in $G$. We show that $S$ contains at least three vertices from every $X$-copy and $Y$-copy in $G$. First we consider an $X$-copy in $G$, and let the vertices in this $X$-copy be named as in Figure~\ref{f:Xcopy}. For notational convenience, we simply call this subgraph $X$. We show that $|S \cap V(X)| \ge 3$.

If $\{a_1, a_2\} \cap S = \varnothing$, then in order to dominate the vertex $b_i$, we note that either $b_i \in S$ or $c_i \in S$. Thus,  $|\{b_i, c_i\} \cap S| = 1$  for all $i \in [3]$, implying that $|S \cap V(X)| \ge 3$, as desired. Hence we may assume that $|\{a_1, a_2\} \cap S| \ge 1$, for otherwise the desired result follows. Renaming $a_1$ and $a_2$ if necessary, we may further assume that $a_1 \in S$. Since $S$ is an independent set, we note that $\{b_1, b_2, b_3\} \cap S = \varnothing$, implying that $a_2 \in S$. We show that the set $S$ contains at least one vertex from the set $\{c_1, c_2, c_3\}$. Suppose, to the contrary, that $\{c_1, c_2, c_3\} \cap S = \varnothing$. By the construction, $c_i$ has exactly one neighbor $c'_i$ in $G \setminus V(X)$ for $i \in [2]$. In order to dominate the vertices $c_1$ and $c_2$, our earlier observations imply that $c'_1,c'_2 \in S$. However by construction, the vertices $c'_1$ and $c'_2$  are adjacent, implying that the set $S$ contains two adjacent vertices, contradicting the fact that $S$ is an independent set. Hence, $\{c_1, c_2, c_3\} \cap S \ne \varnothing$, implying that $|S \cap V(X)| \ge 3$, as desired.

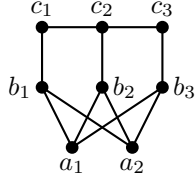
\begin{figure}[htb]
\begin{center}
\begin{tikzpicture}[scale=.8,style=thick,x=1cm,y=1cm]
\def\vr{2.5pt} 
\path (0.5,0) coordinate (a1);
\path (1.5,0) coordinate (a2);
\path (0,1) coordinate (a3);
\path (0,2) coordinate (a6);
\path (1,1) coordinate (a4);
\path (1,2) coordinate (a7);
\path (2,1) coordinate (a5);
\path (2,2) coordinate (a8);
%
\draw (a1) -- (a3);
\draw (a1) -- (a4);
\draw (a1) -- (a5);
\draw (a2) -- (a3);
\draw (a2) -- (a4);
\draw (a2) -- (a5);
\draw (a6) -- (a3);
\draw (a7) -- (a4);
\draw (a8) -- (a5);
\draw (a7) -- (a6);
\draw (a7) -- (a8);
\draw (a1) [fill=black] circle (\vr);
\draw (a2) [fill=black] circle (\vr);
\draw (a3) [fill=black] circle (\vr);
\draw (a4) [fill=black] circle (\vr);
\draw (a5) [fill=black] circle (\vr);
\draw (a6) [fill=black] circle (\vr);
\draw (a7) [fill=black] circle (\vr);
\draw (a8) [fill=black] circle (\vr);
\draw[anchor = north] (a1) node {{\small $a_1$}};
\draw[anchor = north] (a2) node {{\small $a_2$}};
\draw[anchor = east] (a3) node {{\small $b_1$}};
\draw[anchor = west] (a4) node {{\small $b_2$}};
\draw[anchor = west] (a5) node {{\small $b_3$}};
\draw[anchor = south] (a6) node {{\small $c_1$}};
\draw[anchor = south] (a7) node {{\small $c_2$}};
\draw[anchor = south] (a8) node {{\small $c_3$}};
\end{tikzpicture}
\end{center}
\vskip -0.7 cm
\caption{A $Y$-copy in $G$} \label{f:Ycopy}
\end{figure}

Next we consider a $Y$-copy in $G$, and let the vertices in this $Y$-copy be named as in Figure~\ref{f:Ycopy}. For notational convenience, we simply call this subgraph $Y$. We show that $|S \cap V(Y)| \ge 3$.

If $\{a_1, a_2\} \cap S = \varnothing$, then as observed earlier, $|\{b_i, c_i\} \cap S| = 1$ for all $i \in [3]$, implying that $|S \cap V(X)| \ge 3$, as desired. Hence we may assume that $|\{a_1, a_2\} \cap S| \ge 1$, for otherwise the desired result follows. Further we may assume that $a_1 \in S$. As observed earlier, this implies that $\{b_1, b_2, b_3\} \cap S = \varnothing$ and $a_2 \in S$. In order to dominate the vertex $c_2$, the set $S$ contains at least one of the vertices $c_1$, $c_2$ and $c_3$. Thus, $|\{c_1, c_2, c_3\} \cap S| \ge 1$, and so $|S \cap V(Y)| \ge 3$, as desired. This completes the proof of Theorem~\ref{t:main1}.~\QED

\section{Proof of Theorem~\ref{t:main2}}
\label{S:main2}

In this section, we present a proof of Theorem~\ref{t:main2}. First we prove that the independent domination number of a subcubic graph with no isolated vertex is at most one-half the order of the graph.

\begin{theorem}
\label{t:bound}
If $G$ is a subcubic graph of order~$n$ with no isolated vertex, then $i(G) \le \frac{1}{2}n$.
\end{theorem}
\proof By linearity, the independent domination number of a graph is the sum of the independent domination numbers of its components. Hence it suffices for us to prove the bound for connected graphs; that is, we prove that if $G$ is a connected subcubic graph of order~$n \ge 2$, then $i(G) \le \frac{1}{2}n$. We proceed by induction on the order~$n \ge 2$. If $n = 2$, then $G = K_2$ and $i(G) = 1 = \frac{1}{2}\times 2$. If $n = 3$, then $G = K_3$ or $G = P_3$, and in both cases, $i(G) = 1 < \frac{1}{2} \times 3$. This establishes the base cases. Let $n \ge 4$ and assume that if $G'$ is a connected subcubic graph of order~$n'$ where $2 \le n' < n$, then $i(G') \le \frac{1}{2}n'$. Let $G$ be a connected subcubic graph of order~$n$. If $G$ is a bipartite graph, then the desired bound follows from Proposition~\ref{p:bipartite}. Hence we may assume that $G$ contains an odd cycle $C$.

First, assume that there exists a vertex $u$ on the cycle $C$ with a leaf neighbor, say $w$, and consider the graph $H = G - \{u,w\}$. Since the two neighbors of $u$ on the cycle $C$ are connected in $H$ by the path $C - u$, the graph $H$ is a connected subcubic graph. Since $n \ge 4$, we note that $|V(H)| = n - 2 \ge 2$. Applying the induction to $H$, we have $i(H) \le \frac{1}{2}|V(H)| = \frac{1}{2}(n-2)$. Every minimum ID-set in $H$ can be extended to an ID-set of $G$ by adding to it the vertex~$w$, implying that $i(G) \le i(H) + 1 \le \frac{1}{2}n$. Hence we may assume that no vertex on the cycle $C$ has a leaf neighbor.

Next, assume there are two consecutive vertices, say $u$ and $w$, on the cycle $C$ both of degree~$2$ in $G$. We now consider the graph $H = G - \{u,w\}$. We note that the graph $H$ is a connected subcubic graph. Further since $n \ge 4$, we note that $|V(H)| = n - 2 \ge 2$.  Applying the induction to $H$, we have $i(H) \le \frac{1}{2}|V(H)| = \frac{1}{2}(n-2)$. Let $S'$ be a minimum ID-set of $H$, and let $u'$ be the neighbor of $u$ different from~$w$ and let $w'$ be the neighbor of $w$ different from~$u$. (Possibly, $u' = w'$.) If $u' \notin S'$, then let $S = S' \cup \{u\}$. If $u' \in S$ and $w' \notin S$, then let $S = S' \cup \{w\}$. If $u' \in S$ and $w' \in S$, then let $S = S'$. In all cases, $S$ is an ID-set of $G$, and so $i(G) \le |S| \le |S'| + 1 = i(H) + 1 \le \frac{1}{2}n$. Hence we may assume that no two consecutive vertices on the cycle $C$ both have degree~$2$ in $G$.

Let $u$ and $v$ be two arbitrary consecutive (adjacent) vertices on the cycle $C$. Suppose that there exists a vertex $w$ of degree~$2$ adjacent to both $u$ and~$v$. In this case, we consider the connected subcubic graph $H = G - \{u,v,w\}$. If $|V(H)| = 1$, then $n = 4$ and $G \cong K_4 - e$ where $e$ is the missing edge of the complete graph $K_4$. In this case, $i(G) = 1 < \frac{1}{2}\times4$. Hence we may assume that $|V(H)| \ge 2$. Applying the induction to $H$ we have $i(H) \le \frac{1}{2}|V(H)| = \frac{1}{2}(n-3)$. Every minimum ID-set of $H$ can be extended to an ID-set of $G$ by adding to it the vertex~$w$, implying that $i(G) \le i(H) + 1 < \frac{1}{2}n$. Hence we may assume that there is no vertex of degree~$2$ adjacent to both $u$ and~$v$.

We now consider the subcubic graph $H = G - \{u,v\}$. With our assumptions, we note that $H$ has at most three components, each of which has order at least~$2$. Let $H_1, \ldots, H_t$ be the components of $H$, and so $t \le 3$. Let $S_i$ be a minimum ID-set of $H_i$ for $i \in [t]$. By the inductive hypothesis, $|S_i| \le \frac{1}{2}|V(H_i)|$ for $i \in [t]$. Let
\[
S' = \bigcup_{i=1}^t S_i.
\]
If $u$ has no neighbor in $S'$, then let $S = S' \cup \{u\}$. If $u$ has a neighbor in $S'$ and $v$ has no neighbor in $S'$, then let $S = S' \cup \{v\}$. If both $u$ and $v$ have a neighbor in $S'$, then let $S = S'$. In all three cases, the set $S$ is an ID-set of $G$, and so $i(G) \le |S| \le |S'| + 1 \le \frac{1}{2}|V(H)| + 1 = \frac{1}{2}n$. This completes the proof of Theorem~\ref{t:bound}.~\QED

\medskip
We are now in a position to present a proof of Theorem~\ref{t:main2}. Recall its statement.

\noindent \textbf{Theorem~\ref{t:main2}}. \emph{If $G$ is a subcubic graph of order~$n$ with no isolated vertex, then $i(G) \le \frac{1}{2}n$. Further, if $G$ is connected, then equality holds if and only if the following holds. \\ [-28pt]
\begin{enumerate}
\item $G \in \{G_1,G_2,G_3,G_4,G_5\}$.
\item $n = 2k$ for some $k \ge 1$ and $G = \coro(P_k)$.
\item $n = 2k$ for some $k \ge 3$ and $G = \coro(C_k)$.
\end{enumerate}
}

\noindent
\proof The upper bound $i(G) \le \frac{1}{2}n$ is a restatement of Theorem~\ref{t:bound}. If $G$ is a connected subcubic graph of order~$n$ that satisfies (a), (b) or (c) in the statement of the theorem, then it is a simple exercise to check that $i(G) = \frac{1}{2}n$. Hence it suffices to prove that if $G$ is a connected subcubic graph of order~$n \ge 2$ satisfying $i(G) = \frac{1}{2}n$, then (a), (b) or (c) in the statement of the theorem hold. We proceed by induction on the order~$n \ge 2$.  We note that $n = 2i(G)$ is even since $i(G)$ is an integer. If $n = 2$, then $G = K_2 = \coro(P_1)$. Suppose that $n = 4$. If $\Delta(G) = 3$, then $i(G) = 1 < \frac{1}{2}\times4$, a contradiction. Hence, $\Delta(G) = 2$, and so either $G = P_4 = \coro(P_2)$ or $G = K_{2,2} = G_1$. This establishes the base cases. Let $n \ge 6$ be even and assume that if $G'$ is a connected subcubic graph of even order~$n'$ where $2 \le n' < n$ satisfying $i(G') = \frac{1}{2}n'$, then (a), (b) or (c) in the statement of the theorem hold. Let $G$ be a connected subcubic graph of order~$n$ satisfying $i(G) = \frac{1}{2}n$. We proceed further with two claims. Recall that $G_2 = K_{3,3}$.

\begin{claim}
\label{c:claim1}
If the graph $G$ contains no support vertex, then $G = G_2$.
\end{claim}
\proof Assume that the graph $G$ contains no support vertex. Thus, every vertex of $G$ has degree at least~$2$ and degree at most~$3$. If $\Delta(G) = 2$, then $G$ is a cycle $C_n$ where $n \ge 6$, and so $i(G) = i(C_n) = \lceil \frac{1}{3}n \rceil  < \frac{1}{2}n$, a contradiction. Hence, $\Delta(G) = 3$. Let $v$ be an arbitrary vertex of degree~$3$ in $G$ and let $N_G(v) = \{x,y,z\}$. We now consider the graph $H = G \setminus N_G[v]$. Suppose that $H$ has no isolated vertex. Let $H_1, \ldots, H_t$ be the components of $H$ and let $S_i$ be a minimum ID-set of $H_i$ for $i \in [t]$. By Theorem~\ref{t:bound}, $|S_i| = i(H_i) \le \frac{1}{2}|V(H_i)|$ for $i \in [t]$. Let
\[
S' = \bigcup_{i=1}^t S_i \hspace*{0.5cm} \mbox{and} \hspace*{0.5cm} S = S' \cup \{v\}.
\]

The set $S$ is an ID-set of $G$, implying that
\[
\begin{array}{lcl}
i(G) \le |S| = 1 + |S'| & = & \displaystyle{ 1 + \sum_{i=1}^t |S_i| } \1 \\
& \le & \displaystyle{ 1 + \sum_{i=1}^t \frac{1}{2}|V(H_i)| } \2 \\
& = & \displaystyle{ 1 + \frac{1}{2}|V(H)| } \2 \\
& = & \displaystyle{ 1 + \frac{1}{2}(n-4) } \2 \\
& < & \displaystyle{ \frac{1}{2}n },
\end{array}
\]
a contradiction. Hence, $H$ contains at least one isolated vertex. If $H$ contains at least four isolated vertices, then since $\delta(G) \ge 2$ each such isolated vertex in $H$ has at least two neighbors in $G$ belonging to the set $\{x,y,z\}$, implying by the Pigeonhole Principle that at least one of the vertices $x$, $y$ and $z$ has degree at least~$4$ in $G$, a contradiction. Therefore, $H$ contains at most three isolated vertices.

We show that $H$ contains at most two isolated vertices. Suppose, to the contrary, that $H$ contains three isolated vertices. Since $\delta(G) \geq 2$ and $\Delta(G) = 3$, the graph $G$ is now determined. In this case, $n = 7$ and this contradicts the fact that $n$ is even. Hence, $H$ contains at most two isolated vertices.
	
Next we show that $H$ contains exactly two isolated vertices. Suppose, to the contrary, that $H$ contains exactly one isolated vertex, say $u$. In this case, we consider the graph $H' = H \setminus \{u\}$. Since $n \ge 6$ is even and $H'$ contains no isolated vertex, every component of $H'$ has order at least~$2$. Applying Theorem~\ref{t:bound} to $H'$, we have $i(H') \le \frac{1}{2}|V(H')| = \frac{1}{2}(n-5)$. Since $n$ is even, this implies that $i(H') \le \frac{1}{2}(n-6)$. A minimum ID-set of $H'$ can be extended to an ID-set of $G$ by adding to it the vertices $u$ and $v$, implying that $i(G) \le i(H') + 2 < \frac{1}{2}n$, a contradiction. Hence, $H$ contains exactly two isolated vertices.

Let $u$ and $w$ be the two isolated vertices of $H$. Each of $u$ and $w$ has either two or three neighbors in $G$ that belong to the set $\{x,y,z\}$, implying that $u$ and $w$ have at least one common neighbor.

Suppose that $u$ and $w$ have exactly one common neighbor. Renaming the neighbors of $v$ if necessary, we may assume that $N_G(u) = \{x,y\}$ and $N_G(w) = \{y,z\}$. In particular, $y$ is the common neighbor of $u$ and $w$. Let $H'=H-\{u,w\}$. We note that $H'$ has no isolated vertex. Applying Theorem~\ref{t:bound}, $i(H') \le \frac{1}{2} |V(H')| = \frac{1}{2}(n-6)$. Let $S$ be a minimum ID-set of $H'$. If $N(z) \cap S=\varnothing$, then $S\cup\{u,z\}$ is an ID-set of G, a contradiction. If $N(x) \cap S = \varnothing$, then similarly we get a contradiction. Now, assume that $N(z) \cap S \neq \varnothing$ and $N(x) \cap S \neq \varnothing$. In this case, $S \cup {y}$ is an ID-set of $G$, a contradiction. Hence, $u$ and $w$ have at least two common neighbors.

Suppose that $u$ and $w$ have exactly two common neighbors. Renaming neighbors of $v$ if necessary, we may assume in this case that $\{x,y\} = N_G(u)\cap N_G(w)$. By assumption, $z$ is adjacent to at most one of $u$ and $w$. Renaming $u$ and $w$, we may assume that $z$ is not adjacent to $w$. If $n = 6$, then $\{w,z\}$ is an ID-set of $G$, implying that $i(G) = 2 < \frac{1}{2}\times 6$, a contradiction. Hence, $n \ge 8$. We now consider the connected subcubic graph $H' = G - \{u,v,w,x,y\}$. Applying Theorem~\ref{t:bound} to $H'$, we have $i(H') \le \frac{1}{2}|V(H')| = \frac{1}{2}(n-5)$. Since $n$ is even, this implies that $i(H') \le \frac{1}{2}(n-6)$. A minimum ID-set of $H'$ can be extended to an ID-set of $G$ by adding to it the vertices $x$ and $y$, implying that $i(G) \le i(H') + 2 < \frac{1}{2}n$, a contradiction.

Hence, the vertices $u$ and $w$ have three common neighbors. The graph $G$ is now determined, and $G = K_{3,3} = G_2$. This completes the proof of the claim.~\smallqed

\medskip
By Claim~\ref{c:claim1}, we may assume that the graph $G$ contains at least one support vertex, for otherwise $G = G_2$ and the desired result follows. Since $n \ge 6$, we note that every support vertex of $G$ has at most two leaf neighbors. Recall that $G_4$ is the double star $S(2,2)$ shown in Figure~\ref{f:special}(b).

\begin{claim}
\label{c:claim2}
If the graph $G$ contains a support vertex with two leaf neighbors, then $G = G_4$.
\end{claim}
\proof Suppose that $G$ contains a support vertex $v$ with two leaf neighbors, say $u$ and $w$. Let $x$ be the third neighbor of $v$. Since $n \ge 6$, we note that $d_G(x) \ge 2$. We show that $x$ is a support vertex. Suppose, to the contrary, that $x$ is not a support vertex. In this case, we consider the subcubic graph $H = G - N_G[v] = G - \{u,v,w,x\}$. Since $x$ is not a support vertex in $G$, every component of $H$ has order at least~$2$. Applying Theorem~\ref{t:bound} to $H$, we have $i(H) \le \frac{1}{2}|V(H)| = \frac{1}{2}(n-4)$. A minimum ID-set of $H$ can be extended to an ID-set of $G$ by adding to it the vertex $v$, implying that $i(G) \le i(H) + 1 < \frac{1}{2}n$, a contradiction. Hence, $x$ is a support vertex.

Next, we show that $x$ has two leaf neighbors. Suppose, to the contrary, that $x$ has exactly one leaf neighbor, say $y$. Since $n \ge 6$, we note that in this case the vertex $x$ has degree~$3$. We consider the connected subcubic graph $H = G - \{u,v,w,x,y\}$. We note that $H$ has order at least~$2$. Applying Theorem~\ref{t:bound} to $H$, we have $i(H) \le \frac{1}{2}|V(H)| = \frac{1}{2}(n-5)$. Since $n$ is even, this implies that $i(H) \le \frac{1}{2}(n-6)$. A minimum ID-set of $H$ can be extended to an ID-set of $G$ by adding to it the vertices $v$ and $y$, implying that $i(G) \le i(H) + 2 < \frac{1}{2}n$, a contradiction. Hence, $x$ has exactly two leaf neighbors; that is, $G = G_4$.~\smallqed

\medskip
By Claim~\ref{c:claim2}, we may assume that every support of $G$ has exactly one leaf neighbor, for otherwise $G = G_4$ and the desired result follows. Among all support vertices of $G$, let $v$ be chosen so that the following holds, where $u$ is the leaf neighbor of $v$. \1 \\
\hspace*{1cm} (1) The degree, $d_G(v)$, of $v$ is a minimum. \\
\hspace*{1cm} (2) Subject to (1), the number of components of $G - \{u,v\}$ is a minimum.

\noindent
We note that either $d_G(v) = 2$ or $d_G(v) = 3$. Further, we note that either $G - \{u,v\}$ is connected or has two components. Let $H = G - \{u,v\}$. Each component of $H$ contains a neighbor of $v$. Since $u$ is the only leaf neighbor of $v$, the graph $H$ has no isolated vertex, and so each component of $H$ has order at least~$2$. Applying Theorem~\ref{t:bound} to $H$, we have $i(H) \le \frac{1}{2}|V(H)| = \frac{1}{2}(n-2)$. A minimum ID-set of $H$ can be extended to an ID-set of $G$ by adding to it the vertex $u$, implying that $\frac{1}{2}n = i(G) \le i(H) + 1 \le \frac{1}{2}n$. Hence we must have equality throughout this inequality chain, implying that $i(H) = \frac{1}{2}|V(H)|$ and that every component $H'$ of $H$ satisfies $i(H') = \frac{1}{2}|V(H')|$. Applying the inductive hypothesis to each component $H'$ of $H$, the component $H'$ satisfies (a), (b) or (c) in the statement of the theorem. Recall that $G_3$ and $G_5$ are the graphs shown in Figure~\ref{f:special}(a) and~\ref{f:special}(b), respectively.

\begin{claim}
\label{c:claim3}
The graph $H$ is connected.
\end{claim}
\proof Suppose, to the contrary, that $H$ is disconnected. Thus, $H$ has two components, say $H_1$ and $H_2$. In particular, this implies that $d_G(v) = 3$. Let $v_i$ be the neighbor of $v$ that belongs to $H_i$ for $i \in [2]$. Let $H_i$ have order~$n_i$ for $i \in [2]$. As observed earlier, $n_i \ge 2$ and $i(H_i) = \frac{1}{2}n_i$ for $i \in [2]$. Further, $H_i$ satisfies (a), (b) or (c) in the statement of the theorem for $i \in [2]$.

\begin{subclaim}
\label{c:claim3.1}
$H_i \notin \{G_1,G_2,G_3,G_4,G_5\}$ for $i \in [2]$.
\end{subclaim}
\proof Suppose, to the contrary, that $H_1 \in \{G_1,G_2,G_3,G_4,G_5\}$. We note that $H_1 \ne G_2 = K_{3,3}$ since the vertex $v_1$ has degree at most~$2$ in $H_1$. Thus, $H_1 = G_1$, in which case $n_1 = 4$, or $H_1 \in \{G_3,G_4,G_5\}$, in which case $n_1 = 6$. Further we note that $H_1 - v_1$ is a connected graph of odd order $n_1 - 1 \ge 3$, implying by Theorem~\ref{t:bound} that $i(H_1 - v_1) \le \frac{1}{2}(n_1 - 2)$. Let $S_1$ be a minimum ID-set of $H_1$. We note that the set $S_1$ contains no neighbor of $v_1$. We now consider the connected subcubic graph $G' = G - (V(H_1) \setminus \{v_1\})$. Let $G'$ has order~$n'$. Since $n' = n - n_1 + 1$ is odd, Theorem~\ref{t:bound} implies that $i(G') \le \frac{1}{2}(n' - 1) = \frac{1}{2}(n - n_1)$. If $S'$ is an ID-set of $G'$ of minimum cardinality, then $S' \cup S_1$ is an ID-set of $G$, implying that
\[
i(G) \le |S_1| + |S'| \le \frac{1}{2}(n_1 - 2) + \frac{1}{2}(n - n_1) < \frac{1}{2}n,
\]
a contradiction.~\smallqed

\medskip
By Claim~\ref{c:claim3.1} and our earlier observations, $H_i$ satisfies (b) or (c) in the statement of the theorem for $i \in [2]$. Thus, $n_i = 2k_i$ and $H_i = \coro(P_{k_i})$ for some $k_i \ge 1$ or $H_i = \coro(C_{k_i})$ for some $k_i \ge 3$ and $i \in [2]$. Recall that among all support vertices of $G$, the vertex $v$ was chosen to have minimum degree. This implies that if $H_i = \coro(P_{k_i})$, then $k_i \ge 2$ for $i \in [2]$, for otherwise if $H_i = \coro(P_1) = P_2$, then the vertex $v_i$ would be a support vertex of $G$ of degree~$2$, a contradiction. In particular, we note that $n_i \ge 4$ for $i \in [2]$. If $H_1 = \coro(P_{k_1})$ for some $k_1 \ge 2$, then at least one of the two support vertices of degree~$2$ in $H$ is a support vertex of degree~$2$ in $G$, contradicting our choice of the support vertex~$v$. If $H_1 = \coro(C_{k_1})$ for some $k_1 \ge 3$, then at least one support vertex (of degree~$3$) in $H$ is a support vertex in $G$. However, the removal of such a support vertex and its leaf neighbor in $G$ produces a connected graph, once again contradicting our choice of the support vertex~$v$. This completes the proof of Claim~\ref{c:claim3}.~\smallqed

\medskip
By Claim~\ref{c:claim3}, the graph $H$ is connected. Let $H$ have order~$n'$, and so $n' = n - 2$. As observed earlier, $H$ satisfies (a), (b) or (c) in the statement of the theorem. Thus, $H \in \{G_1,G_2,G_3,G_4,G_5\}$ or $n' = 2k'$ for some $k' \ge 1$ and $H = \coro(P_{k'})$ or $n' = 2k'$ for some $k' \ge 3$ and $H = \coro(C_{k'})$.

\begin{claim}
\label{c:claim4}
If $H \in \{G_1,G_2,G_3,G_4,G_5\}$, then $G = G_3$.
\end{claim}
\proof Suppose that $H \in \{G_1,G_2,G_3,G_4,G_5\}$. We consider each possibility in turn. Suppose that $H = G_1 = C_4$, and so $n = 6$. Let $H$ be the cycle $C \colon w_1w_2w_3w_4w_1$, where $vw_1$ is an edge of $G$. If $vw_3$ is not an edge of $G$, then $\{v,w_3\}$ is an ID-set of $G$, and so $i(G) = 2 < \frac{1}{2}\times 6$, a contradiction. Hence, $vw_3$ is an edge of $G$, implying that $G = G_3$.

We note that the vertex $v$ has one or two neighbors in $H$, and each neighbor of $v$ in $H$ has degree at most~$2$ in $H$, implying that $H \ne G_2 \cong K_{3,3}$.

Suppose that $H = G_3$, and so $n = 8$. Let $a_1$ and $a_2$ be the two vertices of $H$ with three common neighbors, say $b_1$, $b_2$ and $b_3$, where $b_3$ has degree~$3$ in $H$. Let $w$ be the leaf neighbor of $b_3$ in $H$. If $vw \in E(G)$, then let $S = \{a_1,a_2,v\}$. If $vw \notin E(G)$ and $v$ is adjacent to both $b_1$ and $b_2$, then let $S = \{b_3,v\}$. If $vw \notin E(G)$ and $v$ is adjacent to exactly one of $b_1$ and $b_2$, say to $b_1$, then let $S = \{b_2,b_3,v\}$. In all three cases, the set $S$ is an ID-set of $G$ and $|S| \le 3$. Thus, $i(G) \le 3 < \frac{1}{2} \times 8$, a contradiction.

Suppose that $H = G_4$, and so $n = 8$. Let $x$ and $y$ be the two central vertices of the double star $H$. By our earlier assumptions, every support vertex of $G$ has exactly one leaf neighbor. Hence, the vertex $v$ is adjacent in $G$ to a leaf neighbor in $H$ of $x$ and a leaf neighbor in $H$ of $y$. Thus if $x'$ be the leaf neighbor of $x$ in $H$ that is not adjacent to $v$ in $G$, then the set $\{v,x',y\}$ is an ID-set of $G$, and so $i(G) \le 3 < \frac{1}{2}\times 8$, a contradiction.

Suppose that $H = G_5$, and so $n = 8$. Thus, $H$ is obtained from a path $a_1a_2a_3a_4a_5a_6$ by adding the edge $a_2a_5$. Suppose that $v$ is adjacent to $a_1$ or $a_6$, say to $a_1$. If $v$ is not adjacent to $a_3$, then let $S = \{v,a_3,a_5\}$. If $v$ is adjacent to $a_3$, then let $S = \{v,a_5\}$. In both cases, the set $S$ is an ID-set of $G$ and $|S| \le 3$. Thus, $i(w1G) \le 3 < \frac{1}{2} \times 8$, a contradiction. Hence, $v$ is adjacent to neither $a_1$ nor $a_6$. Thus, the only possible neighbors of $v$ in $H$ are $a_3$ or $a_4$. By symmetry, we may assume that $va_3 \in E(G)$. Thus, $\{v,a_1,a_5\}$ is an ID-set of $G$, and so $i(G) \le 3 < \frac{1}{2} \times 8$, a contradiction. This completes the proof of Claim~\ref{c:claim4}.~\smallqed

\medskip
Let $n' = |V(H)|$, and so $n' = n - 2$. Recall that $n \ge 6$, and so $n' \ge 4$. By Claim~\ref{c:claim4}, we may assume that $H \notin \{G_1,G_2,G_3,G_4,G_5\}$, for otherwise $G = G_3$, and the desired result follows. Hence $n' = 2k'$ and $H = \coro(P_{k'})$ for some $k' \ge 2$ or  $H = \coro(C_{k'})$ for some $k' \ge 3$.

\begin{claim}
\label{c:claim5}
$H = \coro(P_{k'})$ for some $k' \ge 2$.
\end{claim}
\proof Suppose that $H = \coro(C_{k'})$ for some $k' \ge 3$. Thus, $n' = 2k'$ and $n = 2k' + 2$. Let $H$ be the corona of the cycle $C \colon v_1v_2 \ldots v_{k'}v_1$, and let $u_i$ be the resulting leaf neighbor of $v_i$ in $H$ for $i \in [k']$. Since $G$ is a subcubic graph, we note that the only possible neighbors of $v$ that belong to $H$ are the leaves of $H$. We now consider the connected subcubic graph $G' = G - N_G[v]$ of order at least~$2$. Applying Theorem~\ref{t:bound} to the graph $G'$, we have $i(G') \le \frac{1}{2}|V(G')| \le \frac{1}{2}(n-3)$. Since $n$ is even, this implies that $i(G') \le \frac{1}{2}(n-4)$. A minimum ID-set of $G'$ can be extended to an ID-set of $G$ by adding to it the vertex $v$, implying that $i(G) \le i(G') + 1 < \frac{1}{2}n$, a contradiction.~\smallqed

\medskip
By Claim~\ref{c:claim5}, $H = \coro(P_{k'})$ for some $k' \ge 2$. Thus, $n' = 2k'$ and $n = 2k' + 2$. Let $H$ be the corona of the path $P \colon v_1v_2 \ldots v_{k'}$, and let $u_i$ be the resulting leaf neighbor of $v_i$ in $H$ for $i \in [k']$. Since $G$ is a subcubic graph, we note that the only possible neighbors of $v$ that belong to $H$ are the vertices $u_i$ for $i \in [k']$ or the vertices $v_1$ and $v_{k'}$ of degree~$2$ in $H$.

\begin{claim}
\label{c:claim6}
If $d_G(v) = 2$, then $G = \coro(P_k)$ where $k = k'+1$.
\end{claim}
\proof Suppose that $d_G(v) = 2$. Let $w$ be the neighbor of $v$ different from $u$. If $w = u_i$ for some $i \in [k']$, then we consider the connected subcubic graph $G' = G - \{u,v,w\}$ of order at least~$3$. Applying Theorem~\ref{t:bound} to the graph $G'$, we have $i(G') \le \frac{1}{2}|V(G')| = \frac{1}{2}(n-3)$. Since $n$ is even, this implies that $i(G') \le \frac{1}{2}(n-4)$. A minimum ID-set of $G'$ can be extended to an ID-set of $G$ by adding to it the vertex $v$, implying that $i(G) \le i(G') + 1 < \frac{1}{2}n$, a contradiction. Hence, either $w = v_1$ or $w = v_{k'}$. In both cases, $G = \coro(P_k)$ where $k = k'+1$, as desired.~\smallqed

\medskip
By Claim~\ref{c:claim6}, we may assume that $d_G(v) = 3$. Hence, the vertex $v$ has two neighbors in $H$, say $w$ and $x$.
If both neighbors $w$ and $x$ are leaves in $H$, then we consider the connected subcubic graph $G' = G - \{u,v,w,x\}$ of order at least~$2$. Applying Theorem~\ref{t:bound} to the graph $G'$, we have $i(G') \le \frac{1}{2}|V(G')| = \frac{1}{2}(n-4)$. A minimum ID-set of $G'$ can be extended to an ID-set of $G$ by adding to it the vertex $v$, implying that $i(G) \le i(G') + 1 < \frac{1}{2}n$, a contradiction. Hence, renaming $w$ and $x$ if necessary, we may assume that $w = v_1$.

If $x = v_{k'}$, then $G = \coro(C_k)$ where $k = k'+1$, and the desired result follows. Hence, we may assume that $x$ is a leaf of~$H$.

If $x = u_1$, then again we consider the connected subcubic graph $G' = G - \{u,v,w,x\}$, and as before obtain the contradiction $i(G) \le i(G') + |\{v\}| < \frac{1}{2}n$. Hence, $x = u_i$ for some $i \in [k'] \setminus \{1\}$. Suppose that $k' \ge 3$. In this case, we consider the connected subcubic graph $G' = G - \{u,v,w,x,u_1\}$ of order at least~$3$. Applying Theorem~\ref{t:bound} to the graph $G'$, we have $i(G') \le \frac{1}{2}|V(G')| = \frac{1}{2}(n-5)$. Since $n$ is even, this implies that $i(G') \le \frac{1}{2}(n-6)$. A minimum ID-set of $G'$ can be extended to an ID-set of $G$ by adding to it the vertices $u_1$ and $v$, implying that $i(G) \le i(G') + 2 < \frac{1}{2}n$, a contradiction. Hence, $k' = 2$, implying that $G = G_5$, and the desired result follows. This completes the proof of Theorem~\ref{t:main2}.~\smallqed


\medskip
\begin{thebibliography}{99}

\bibitem{AbHe18} G. Abrishami, and M.A. Henning, Independent domination in subcubic graphs  of girth at least six, \textit{Discrete Math.} \textbf{341} (2018), 155--164.

\bibitem{BrHe19} C. Brause and M.A. Henning, Independent domination in bipartite cubic graphs, \textit{Graphs Combin} \textbf{35}(4) (2019), 881--919.

\bibitem{DoHeMoSo15} P. Dorbec, M.A. Henning, M. Montassier, and J. Southey, Independent domination in cubic graphs, \textit{J. Graph Theory} \textbf{80}(4) (2015), 329--349.

\bibitem{Fa} O. Favaron, A bound on the independent domination number of a tree, \textit{Vishwa Internat. J. Graph Theory} \textbf{1} (1992), 19--27.

\bibitem{Furuya} M. Furuya, K. Ozeki, and A. Sasaki, On the ratio of the domination number and the independent domination number in graphs, \textit{Discrete Appl. Math.} \textbf{178} (2014), 157--159.

\bibitem{GoHe13} W. Goddard and M.A. Henning, Independent domination in graphs: A survey and recent results, \textit{Discrete Math}. \textbf{313} (2013), 839--854.

\bibitem{GoHeLySo12} W. Goddard, M.A. Henning, J. Lyle, and J. Southey, On the independent domination number of regular graphs, \textit{Annals Combin}. \textbf{16} (2012), 719--732.

\bibitem{GoLy12} W. Goddard and J. Lyle, Independent dominating sets in triangle-free graphs, \textit{J. Comb. Optim.} \textbf{23}(1) (2012), 9--20.

\bibitem{Ha95} J.~Haviland, Independent domination in regular graphs, \textit{Discrete Math.} \textbf{143} (1995), 275--280.

\bibitem{Ha07} J.~Haviland, Upper bounds for independent domination in regular graphs, \textit{Discrete Math.} \textbf{307} (2007), 2643--2646.

\bibitem{HeLoRa14} M.A. Henning, C. L\"{o}wenstein, and D. Rautenbach,
    Independent domination in subcubic bipartite graphs of girth at least six, \textit{Discrete Appl. Math.} \textbf{162}  (2014), 399--403.	


\bibitem{HeYe_book} M.A. Henning and A. Yeo, \emph{Total Domination in Graphs (Springer Monographs in Mathematics)} 2013, ISBN: 978-1-4614-6524-9 (Print) 978-1-4614-6525-6 (Online).



\bibitem{Ko93} A.V. Kostochka, The independent domination number of a cubic $3$-connected graph can be much larger than its domination number, \textit{Graphs Combin.} \textbf{9}(3) (1993), 235--237.

\bibitem{LSS99} P.C.B. Lam, W.C. Shiu, and L.~Sun, On independent domination number of regular graphs, \textit{Discrete Math.} \textbf{202} (1999), 135--144.


\bibitem{Ly14}	J. Lyle, A note on independent sets in graphs with large minimum degree and small cliques, \textit{Electr. J. Comb.} \textbf{21}(2) P2.38 (2014).

\bibitem{Ly15}	J. Lyle, A structural approach for independent domination of regular graphs, \textit{Graphs Combin.} \textbf{31}(5) (2015), 1567--1588.

\bibitem{OWe10} S. O and D.B. West, Cubic graphs with large ratio of independent domination number to domination number, \textit{Graphs Combin.} \textbf{32}(2) (2016), 773--776.
	
\bibitem{RaVo13} N.J. Rad and L. Volkmann, A note on the independent domination number in graphs, \textit{Discrete Appl. Math.} \textbf{161} (2013) 3087--3089.

\bibitem{SoHe13} J. Southey and M.A. Henning, Domination versus independent domination in cubic graphs, \textit{Discrete Math.} \textbf{313}(11) (2013), 1212--1220.

\bibitem{WaWe} S. Wang and B. Wei, A note on the independent domination number versus the domination number in bipartite graphs, \textit{Czechoslovak Math. J.} \textbf{67}(2) (2017), 533--536.
	

\end{thebibliography}
\end{document}